\documentclass[final,3p,times,twocolumn]{elsarticle}
\usepackage{amssymb,amscd,amsmath,graphicx}
\biboptions{sort&compress}

\begin{document}

\begin{frontmatter}

\title{Subdiffusive Brownian ratchets rocked by a periodic force}
\author{Igor Goychuk \corref{cor1}
}
\ead{igor.goychuk@physik.uni-augsburg.de}
\ead{www.physik.uni-augsburg.de/$^\sim$igor}
\cortext[cor1]{Tel.:+49-821-598-3235, Fax:+49-821-598-3222}
\address{Institute of Physics, University of Augsburg,
Universit\"atsstr. 1, D-86135 Augsburg, Germany}
\begin{abstract} 
This work puts forward a generalization
of the well-known rocking Markovian Brownian ratchets to the realm of 
antipersistent non-Markovian
subdiffusion in viscoelastic media. 
A periodically forced subdiffusion in a parity-broken 
ratchet potential is considered within 
the non-Markovian
generalized Langevin equation (GLE)  description with a 
power-law memory kernel $\eta(t)\propto t^{-\alpha}$ ($0<\alpha<1$). It is shown 
that  subdiffusive
rectification currents, defined through the mean displacement and subvelocity $v_{\alpha}$, 
$\langle \delta x(t)\rangle\sim v_{\alpha} t^{\alpha}/
\Gamma(1+\alpha)$, emerge asymptotically 
due to the  breaking of the detailed
balance symmetry by driving. The asymptotic exponent is $\alpha$,
the same as for free subdiffusion, $\langle \delta x^2(t)\rangle\propto t^\alpha$.
 However,
a transient to this regime with some time-dependent $\alpha_{\rm eff}(t)$ gradually decaying
in time,
$\alpha\leq \alpha_{\rm eff}(t)\leq 1$, can be very slow
depending on the barrier height and the driving field strength. 
In striking contrast to its normal
diffusion counterpart, the anomalous rectification current is absent asymptotically 
in the limit
of adiabatic driving with frequency $\Omega\to 0$, displaying a
resonance like dependence on the driving frequency. 
However, an anomalous current inversion occurs for a sufficiently fast 
driving, like in the normal diffusion case. In the lowest order of the driving
field, such a rectification current presents a quadratic response effect. 
Beyond perturbation regime it exhibits a broad maximum versus the driving
field strength. Moreover,  anomalous current exhibits a maximum versus 
the potential amplitude.

\end{abstract}

\end{frontmatter}

\section{\label{Intro} Introduction}

\begin{figure}[t]
\centering
\includegraphics[width=8cm]{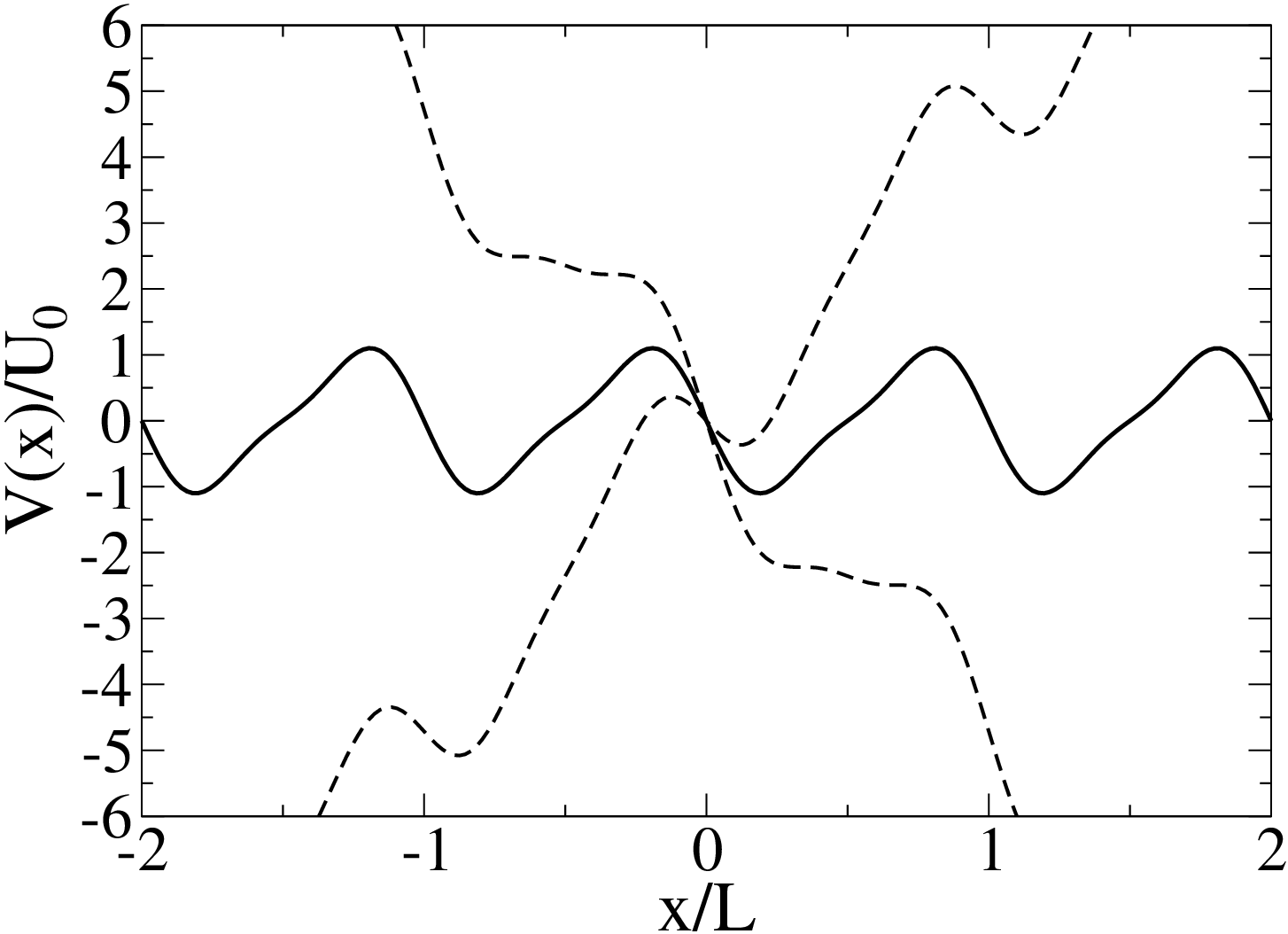}
\caption{Considered ratchet potential, Eq. (\ref{potential}) without
bias and under the critical forward tilt (potential barriers disappear
for the motion to the right), as well as under 
the opposite tilt of the same magnitude. 
Take a note that the tilt in  the opposite
direction does not result in a barrierless total potential. This intuitively
selects  (think about an overdamped particle sliding in the tilted
potential) the right direction for the net current in the case of a
slow time-periodic forcing, symmetric and unbiased on average. For a smaller
rocking force amplitude, the potential barriers in the right sliding  direction are
typically smaller.
} 
\label{Fig1}
\end{figure}

The phenomenon of  directed current in  unbiased on average periodic  potentials,
such as one in Fig. \ref{Fig1}, due
to a violation of the thermal detailed balance symmetry, -- the
so-called ratchet effect \cite{Magnasco,Bartussek,Doering,Ajdari,Astumian,Prost} --  
produced a huge
literature (see, e.g.  Refs. \cite{Reimann,Hanggi09} for
recent reviews and further references).
It was studied at length, inspired in particular by some biological 
molecular motors (e.g. one-headed kinesins C351) \cite{Julicher,Okada,Inoue,Nelson}. 
At the same time,
the fascinating topic of  anomalous diffusion 
\cite{Hughes,Scher,Shlesinger,Bouchaud,Havlin} and, in particular, 
subdiffusion,
where the position variance grows sublinearly in time, i.e. $\langle \delta
x^2(t)\rangle \sim t^{\alpha}$ with $0<\alpha<1$,
gained dramatically in the importance and interest. 
This interest is promoted not in the last line by the
emerging evidence for subdiffusion in 
such complex media as actin filament networks, other protein solutions, interiors
of biological cells 
\cite{Mason,Amblard,Saxton,Qian,Seisenberger,Caspi,Tolic,Weiss1,Banks,
Golding,Mizuno,Szymanski,Pan},  as well as
in the conformational dynamics of protein 
macromolecules \cite{McCammon,Bizzarri,Yang1,Goychuk04,Kneller,Kou,Granek,Luo}.
At the moment, subdiffusion is experimentally
firmly established  in different media 
with different $\alpha$'s in the range of $0.25-0.9$. 
It is thus natural that these two
paradigms-breaking(-and-forming) research lines are currently crossing. 

Whether thermal 
ratchets based on anomalous subdiffusion  are possible 
or not presents a highly nontrivial
challenge, even on the level  of basic and somewhat
oversimplified models. In
particular, a periodically forced continuous time random walk (CTRW) 
subdiffusion and the associated fractional 
Fokker-Planck dynamics \cite{MetzlerPRL,MetzlerKlafter,Sokolov06,PRL07} 
turned out to be not  sensitive to the action of  time-periodic 
fields in the long time limit with an exception of enhancement of
unbiased subdiffusion 
 \cite{Sokolov06,PRL07,Heinsalu09}. 
This fact seems to rule out the very
possibility  of rocking (i.e. fluctuating tilt \cite{Magnasco,Bartussek}) 
subdiffusive ratchets 
based on the CTRW mechanism \cite{Hughes,Scher,Shlesinger,Goychuk06,Heinsalu06}, 
and possibly also on the quenched disorder \cite{Hughes,KlafterSilbey}, 
leaving, however, a door for the flashing (i.e.
pulsating potential \cite{Ajdari,Prost}) subdiffusive 
ratchets based on these mechanisms open \cite{Harms}. 

In biological applications,
the subdiffusion scenario based on the alternative to CTRW, 
fractional Brownian motion (FBM) \cite{Mandelbrot,Feder}
and associated generalized Langevin equation (GLE)  
descriptions \cite{Wang,Kou,Kupferman,Chaudhury,GoychukHanggi07}
can, however, be more relevant \cite{Szymanski,Goychuk09}, 
in particular, because of 
such a subdiffusion is ergodic \cite{Deng,Goychuk09}.
It does not require a quasi-infinite 
 mean residence time
in any  tiny spatial domain, or trap \cite{Goychuk09}, and 
in contrast with the CTRW 
subdiffusion \cite{Barkai} does not suffer
from such non-ergodic features as spontaneous immobilization of some
particles for the time of observation \cite{Heinsalu06}. Given a small number
of functionally specific (sub)diffusing macromolecules, a 
sudden ''non-ergodic'' standstill could
have fatal consequences for the cell functioning.

However, what is the physical origin of GLE subdiffusion? 
In complex environments, such as interior of biological cells densely
stuffed with different macromolecules, including actin filaments, 
either mobile, or building up
the cytoskeleton  networks, static, as in eucaryots, or
dynamically flickering, as in bacteria,  it can
be due to long range negative correlations in the diffusing particle
displacements (and velocity alternations) caused by crowding. 
A particle moving with an instant velocity
$v(t)$ in certain direction invokes besides a  viscous drag also a
quasi-elastic  response of environment opposing the motion.
The total dissipative force can be written in linear approximation
as $F_{v-el}(t)=-\int^t_{-\infty}\eta(t-t')v(t')dt'$
with a memory friction kernel $\eta(t)$. For an oscillatory
motion, $v(t)\sim\hat
v(\omega)\exp(-i\omega t)+c.c$, with frequency $\omega$,  the friction
force becomes frequency-dependent, with amplitude  $\hat
F_{v-el}(\omega)=-\hat \eta (\omega)\hat v(\omega)$,
where $\hat \eta(\omega)=\int_{0}^{\infty}\eta(t)\exp(i\omega t)dt$
is the frequency-dependent friction~\footnote{It is proportional to a frequency-dependent
viscosity, $\hat \zeta(\omega)$, $\hat\eta(\omega)\propto\hat\zeta(\omega)$, 
which was introduced
in the theory of viscoelasticity by A. Gemant \cite{Gemant}
along with the formalism of fractional derivatives.}.  The
subdiffusive behavior can emerge for some special forms of $\eta(t)$
(see below) when  this friction diverges at
zero-frequency, i.e. $\hat \eta(0)=\int_0^{\infty} \eta(t)dt\to
\infty$, or also on some sufficiently long transient time scale,
when $\hat \eta(0)$ is large but finite, which is a more realistic assumption.
Furthermore, the fluctuation-dissipation theorem (FDT) dictates that at
thermal equilibrium  the corresponding energy loss due to dissipation
must always be compensated on  average by an energy gain due to
unbiased (on average) random force of  environment $\xi(t)$.  For  a
particle of mass  $m$ initially localized ($v(t)=0$ for $t<t_0=0$) 
this leads to  GLE description 
\cite{Kubo,Zwanzig,HTB90,ZwanzigBook,WeissBook,NitzanBook},
\begin{equation}\label{GLE} 
m\ddot x+\int_{0}^{t}\eta(t-t')\dot
x(t')dt'+ \frac{\partial V(x,t)}{\partial x} =\xi(t) \;,
\end{equation} 
where the memory kernel and
the autocorrelation function of noise are related by the
fluctuation-dissipation relation (FDR) \cite{Kubo}
\begin{equation}\label{FDR}
\langle \xi(t)\xi(t')\rangle=k_BT\eta(|t-t'|) \; 
\end{equation} 
at temperature $T$.
The thermal random force $\xi(t)$ has necessarily 
Gaussian statistics within the {\it linear} friction approximation considered here
\cite{ReimannChemPhys}, i.e. within {\it linear} FDT,  but {\it not}
necessarily otherwise \cite{Dubkov}. 

In the following,  $V(x,t)=U(x)-x f(t)$, includes some
periodic spatial potential $U(x)=U(x+L)$ with period $L$. It is 
modulated in time by an  
external force $f(t)$. In the numerical simulations below it will be considered 
harmonic, $f(t)=A\cos(\Omega t)$,
with amplitude $A$ and angular frequency $\Omega$. 
 Apart from above phenomenological
justification, the GLE description can be derived from the Hamiltonian
dynamics \cite{Bogolyubov,Kubo}, i.e. from the first principles. All
this makes the GLE approach to anomalous diffusion and relaxation
processes ever more attractive. For spatially and temporally unbiased
(on average) force fields $f(x,t)=-\partial V(x,t)/\partial x$, a
rectification current (i.e. ratchet effect) can emerge if only the
symmetry of detailed balance is destroyed by an external driving \cite{Reimann} which
provides also an energy supply to drive the rectified dissipative motion
in certain direction. 

A popular model of viscoelasticity with 
\begin{eqnarray}\label{Gemant}
\eta(t)=\eta_{\alpha}t^{-\alpha}/\Gamma(1-\alpha)\;,
\end{eqnarray}
where $0<\alpha<1$
(the factor $\Gamma(1-\alpha)$ is   to relate
our description with others\footnote{Then the frictional term of GLE can
be abbreviated as $\eta_{\alpha}\frac{d^\alpha x(t)}{dt^\alpha}$ making use of
the definition of fractional Caputo derivative of the fractional order $\alpha$ 
\cite{Gorenflo}. 
For this particular kernel, 
the corresponding
GLE is named sometimes \it{fractional}.}),
was introduced by Gemant \footnote{More precisely, he introduced 
a more general model for the complex viscosity
which in a particular case  yields
$\hat \zeta(\omega)=\zeta_{\alpha}\tau_c^{1-\alpha}/
[1+(-i\omega\tau_c)^{1-\alpha}]$. The corresponding
$\eta(t)\propto \zeta(t)$ matches Eq. (\ref{Gemant}) for $t\ll\tau_c$, 
while for $t\gg\tau_c$ its 
asymptotics is $\eta(t)\propto
t^{\alpha-2}$. In Ref. \cite{Cole}, Cole and Cole
have remarked that 
the choice $\hat \zeta(\omega)=\zeta_{\alpha}/(-i\omega)^{1-\alpha}$
following to Gemant \cite{Gemant}
(our notations are different) leads to their now famous
form of the dielectrical response function. This corresponds to the 
limit $\tau_c\to \infty$.}\cite{Gemant}.
It allows to rationalize
the Cole-Cole dielectric response of harmonically bound, $U(x)=k x^2/2$, 
overdamped particles in the inertialess limit, 
$m\to 0$ \cite{Cole,Goychuk07}. Moreover, this model
corresponds to the so-called sub-Ohmic spectral bath density
$J(\omega)\propto \eta_{\alpha}\omega^\alpha$ in the language of the Hamiltonian 
system-bath description leading to GLE upon integration over
the initially canonically distributed 
bath variables at temperature $T$. The corresponding relation between $\eta(t)$
and $J(\omega)$ is
$\eta(t)=(2/\pi)\int_0^{\infty} d\omega J(\omega)\cos(\omega t)/\omega$ 
\cite{WeissBook}.

Within this model the variance of
free subdiffusion, or subdiffusion biased by a constant force evolves in time 
asymptotically (for $m\to 0$, exactly if to assume initial velocities thermally
distributed) as \cite{Wang,Kupferman}
\begin{eqnarray}
\langle \delta x^2(t)\rangle \sim 2 D_{\alpha}^{(0)} t^\alpha/\Gamma(1+\alpha),
\end{eqnarray}
where $D_{\alpha}^{(0)}$ is the subdiffusion coefficient, which is related to the anomalous
friction coefficient $\eta_{\alpha}$ and temperature $T$ by the generalized
Einstein-Stokes relation, $D_{\alpha}^{(0)}=k_BT/\eta_{\alpha}$.

The strict power law kernel represents, however, 
rather a theoretical abstraction. All
the realistic power law kernels have cutoffs.  A particular functional form with (upper)
incomplete gamma-function, 
\begin{eqnarray}
\eta(t)=\alpha\eta_{\alpha}\tau_c^{-\alpha}
\Gamma(-\alpha,t/\tau_c)/\Gamma(1-\alpha),
\end{eqnarray}
corresponds to the Davidson-Cole
dielectrical susceptibility of the harmonically bound particles 
\cite{Goychuk07} which is also typical 
for complex fluids, gels and glasses. For $t\ll \tau_c$, this kernel coincides
with one in Eq. (\ref{Gemant}). However,
for $t>\tau_c$ it has an exponential cutoff which makes
the zero-frequency friction $\hat \eta(0)=\alpha\eta_{\alpha}\tau_c^{1-\alpha}$
finite. In such  a more realistic case, subdiffusion
occurs on the time scale $\tau_b<t\ll \tau_c$ and turns over into normal
diffusion for $t> \tau_c$. $\tau_c$ can be, however, large enough
(e.g. seconds to minutes, as in the interior of biological cells)
for the subdiffusion-limited reactions to become important, 
in particular because of a finite size of biological 
cells since macromolecules can subdiffuse over the cell volume
within some time less than $\tau_c$. Initially diffusion is ballistic
on the time scale $0<t<\tau_b$ due to inertial effects, 
where ballistic time $\tau_b$ depends on the details of memory kernel and mass $m$.
It physically corresponds to the relaxation time scale of the 
momentum (on this time scale the ballistic superdiffusion
is persistent). 
Interestingly enough, a short-time superdiffusion was already experimentally 
observed in 
viscoelastic fluids \cite{Atak}.

The noise $\xi(t)$ corresponding to the  model in Eq. (\ref{Gemant})  is mathematically 
the fractional Gaussian noise (fGn) by Mandelbrot and van Ness \cite{Mandelbrot}.
It provides an important instance of the so-called $1/f$ noises which encompass
noises with a low-frequency power law feature in their power spectrum, 
$S(\omega)\propto 1/\omega^\gamma, 0<\gamma<2$\cite{Weissman}, here $\gamma=1-\alpha$.
Moreover, in the absence of potential  the considered model reproduces in the 
limit $m\to 0$ the fractional Brownian motion (FBM) with anti-persistence 
\cite{Mandelbrot,Feder}. The motion in periodic potentials
is also antipersistent and subdiffusive \cite{Goychuk09}. 
Moreover, it has been recently noticed 
that for $\alpha<\alpha_c\approx 0.41$ the inertial effects
are generally important \cite{Burov1}. With their inclusion, 
this model explains also 
a generalized Rocard dielectric response \cite{Burov2} of 
harmonically bound particles
($\tau_c\to \infty$) which can account also
for the fast $\beta-$relaxation due to the cage effect. A corresponding 
generalization of the Davidson-Cole succeptibility to include inertial effects
can also be readily given (as well as for any other memory kernel and model
of anomalous dielectric relaxation and ``glassy'' behavior). 

For particular viscoelastic media, the corresponding 
memory kernels $\eta(t)$ can be derived from 
the rheological measurements  \cite{Mason,Amblard,Qian,Mizuno}. 
This is a phenomenological viewpoint. Moreover, one can
derive many memory kernels from a theoretical approach, 
e.g. from the mode coupling theory \cite{ZwanzigBook} or from the 
theory of viscoelasticitity of  semiflexible polymer entangled 
networks \cite{Gittes}.  The later one yields e.g. a power law kernel
with $\alpha=0.75$ which agrees with some experiments  \cite{Amblard,Mizuno}, 
and  possesses also a long-time memory cutoff naturally
related to the polymer network properties.

Moreover, such a GLE subdiffusion
with $\alpha=0.5$
has been used to mimic \cite{Taloni1} a potential-free single-file 
subdiffusion of hard core 
Brownian particles due to a geometric
crowding effect (i.e., a particle cannot move farther than
some typical distance until another, road-blocking particle 
moves) \cite{Jepsen,Levitt,Taloni2}.
The latter one has been experimentally realized and 
observed with colloidal
particles \cite{Wei,Lin}.
Even if the presence of a biasing force acting on {\it all} particles
clearly destroys this correspondence (since all the particles 
drift normally in the same direction),
one can imagine a situation where an external field acts only on
{\it some}, e.g., electrically charged guest particles. 
Then one can expect that this 
correspondence might
hold further and the behavior of the charged tracer particles (e.g. a globular
proteins  in a suspension of neutral 
colloidal particles confined in a quasi-one-dimensional geometry) 
in some electrical potential $V(x,t)$ externally imposed 
can yet be mimicked by GLE in agreement with \cite{Taloni1,Lizana}.
Similar colloidal systems \cite{Wei,Lin} can be considered as 
plausible candidates 
for an experimental realization of subdiffusive ratchets with 
$\alpha=0.5$ which we consider in the following for the purpose
of illustration. The qualitatively same effects
were also found for $\alpha=0.75$ and are expected for other 
values of $\alpha$ as well.

The emergence of subdiffusive ratchet effect is highly nontrivial.    
It has been recently shown numerically for sinusoidal potential 
\cite{Goychuk09} that such a  viscoelastic
subdiffusion is asymptotically not sensitive to the amplitude of the
periodic potential, in agreement with \cite{Chen}
\footnote{The model in Ref. \cite{Chen} is quantum-mechanical and it
includes tunneling effects. In the quantum case, this result is also supported
by the so-called duality transformation mapping between
the dissipative tight-binding
and full potential problems with different 
bath spectral densities
\cite{WeissBook}. However, our results show that this puzzling effect is
of purely classical origin.  It is invoked by the anti-persistent nature of
GLE subdiffusion. }. 
This fact might seem to
rule out the very possibility of such a rocking ratchet as
well.  On the contrary, we show in this work that such subdiffusive 
Brownian ratchets
are possible and moreover they display
a resonance-like dependence on the driving frequency $\Omega$.
The rectification current {\it vanishes} in the limit
of adiabatic rocking, in a clear contrast with the normal diffusion
rocking ratchets, where it is maximal \cite{Reimann,Hanggi09}.
This behavior somewhat resembles pulsing potential ratchets of normal
diffusion. The physical mechanisms are, however, different. In particular, 
our ratchet also exhibits the phenomenon of (sub)current
inversion at sufficiently fast driving, similar to its normal diffusion
counterpart \cite{Bartussek}, whereas the subdiffusion coefficient remains
rather robust and weakly sensitive to the details of potential
and driving, being close to that of free subdiffusion.
With a further increase of driving frequency the rectification
current gradually vanishes.
We shall demonstrate below numerically and explain how this
puzzling 
non-adiabatically rocking, anomalous subdiffusive ratchet mechanism 
operates.

\section{ Theory} 

For arbitrary potentials, including 
 a typical ratchet potential  \cite{Bartussek}
\begin{equation}\label{potential}
 U(x)=-U_0[\sin(2\pi x/L)+(1/4)\sin(4\pi x/L)]
\end{equation}
with period $L$ and height $U_0$, which we consider henceforth,
the exact non-Markovian 
Fokker-Planck equation (NMFPE) which corresponds to the considered GLE model
is not known. The cases of linear and parabolic $U(x)$ present
the only known exceptions
\cite{Adelman,Hanggi1,Hanggi2,Hanggi3,HanggiMojtabai,Hynes} 
which cannot much help in the
context of ratchet problem. 
This, in particular, 
makes a rigorous analytical study of subdiffusion in nontrivial potentials
a highly notrivial problem without exact analytical solutions.

We approach it numerically by embedding 
the non-Markovian subdiffusive dynamics as a Markovian 
stochastic dynamics into the
phase-space of higher dimensionality following to Ref. \cite{Goychuk09}. 
Namely, the
considered power law kernel is approximated by a sum  of $N$-exponentials,
\begin{eqnarray}\label{approx}
\eta(t)=\sum_{i=0}^{N-1}\eta_i\exp(-\nu_i t),
\end{eqnarray}
with $\nu_i=\nu_0/b^i$ and 
$\eta_i=(\eta_{\alpha}/\Gamma(1-\alpha))C_{\alpha}(b)
\nu_0^\alpha/b^{i\alpha}$ scaled hierarchically using a 
dilation parameter $b>1$. In the theory of anomalous relaxation
similar expansions are well-known \cite{Palmer,Hughes}. In the present context, 
the approach corresponds to expansion of 
fractional Gaussian noise into a sum of uncorrelated 
Ornstein-Uhlenbeck (OU) noises, 
$\xi(t)=\sum_{i=0}^{N-1}\zeta_i(t)$, with 
autocorrelation functions, $\langle\zeta_i(t)\zeta_j(t')\rangle=k_BT\eta_i
\delta_{ij}\exp(-\nu_i|t-t'|)$. Then, the GLE (\ref{GLE}) with
the memory kernel in (\ref{approx})  can
be obtained by eliminating the auxiliary variables $u_i(t)$ from
the following Markovian stochastic dynamics in the $D=N+2$ dimensional phase space 
$(x,v,u_0,...,u_{N-1})$:  
\begin{eqnarray}\label{embedding}
\dot x&=& v\;,\nonumber \\
m\dot v & =& -\frac{\partial V(x,t)}{\partial x}+
\sum_{i=0}^{N-1}u_i(t) \;,\nonumber \\
\dot u_i& = &-\eta_i v-\nu_iu_i+\sqrt{2\nu_i\eta_ik_BT}\xi_i(t) \;,
 \end{eqnarray}
where $\xi_i(t)$ are independent unbiased white Gaussian noise sources, 
$\langle \xi_i(t)\xi_j(t')\rangle=\delta_{ij}\delta(t-t')$. To
enforce the FDR (\ref{FDR}) for all times, 
the initial $u_i(0)$ are independently distributed
with the standard deviations $\sigma_i=\sqrt{k_BT\eta_i}$ and 
zero mean \cite{Kupferman,Goychuk09,Siegle}.
The idea of such a Markovian embedding of non-Markovian GLE dynamics is
pretty old \cite{Marchesoni} and the embedding is not unique 
\cite{Kupferman,Siegle}. However, our scheme is the simplest one which serves
the purpose and it leads to an insightful and simple interpretation of the 
anomalous rate processes in terms of slowly fluctuating  rates, with 
the rms amplitude of rate fluctuations which is gradually dying out upon 
increasing the potential barriers. For  sufficiently
high potential barriers (e.g. exceeding $12\;k_BT$ for $\alpha=0.5$ and
about $9\;k_BT$ for $\alpha=0.75$
\cite{Goychuk09}) the rate description for the escape processes is restored.
Then the escape rate becomes excellently described by the celebrated non-Markovian rate theory
\cite{HTB90,Grote,HanggiMojtabai,Pollak}.

Moreover, independently of the FBM connection, the model in Eq. (\ref{embedding})
 can be considered as a
physically plausible viscoelastic model in its own rights 
as any power law dependence 
{\it observed} in the nature can be approximated by a finite sum 
of exponentials and such an expansion is a standard methodology e.g. in
spectroscopy (approximation of spectra by a sum of Lorentzians) 
and in modeling of ion channel kinetics in biology. 
 The inverse of $\nu_0$ corresponds to the fastest  time
scale of the physical noise, or the high-frequency cutoff 
introduced into the (thus approximated) fGn, and
$\nu_{N-1}=\nu_0/b^{N-1}$ corresponds to the low-frequency cutoff.
Such or similar cutoffs are always present in any realistic physical
setup \cite{Weissman}. Even if on the time scale  
$t>b^{N-1}/\nu_0=\tau_c$ the subdiffusion will turn over into the normal diffusion,
this time can be extremally large and practically not reachable,
as in our simulations (or sufficiently large for subdiffusion to become
physically relevant, as in real experiments).
As a rule of thumb, a power law 
extending over
$N$ time-decades can be reasonably well approximated by a sum of about 
$N$ exponentials
choosing the dilation parameter $b=10$. The number $N$ can thus be 
surprisingly small. For example, an experimental power law extending 
over about 5 time decades was nicely fitted to a sum of 6 exponentials in 
Ref. \cite{Sansom} in a more flexible way, i.e. not assuming a
precise scaling \footnote{Of course, fitting 
a power law by a sum of exponentials does not answer the question about
which physics is behind the power law scaling. However,
it offers the way for an economical Markovian embedding of the observed
non-Markovian dynamics. Moreover, the sum of exponentials can reflect
some real hierarchical, tier like structured relaxation with a finite
number of tier levels \cite{Bouchaud}.}.  
So, choosing $\nu_0=10^{3}$ (arbitrary units)
and $N=16$, one can fit with $C_{1/2}(10)=1.3$ 
the power law kernel for $\alpha=0.5$
from $t=10^{-3}$ till $t=10^{11}$, i.e. over $14$th time decades 
\cite{Goychuk09}. The ``error'' (if to think of the model
of fractional Brownian motion with inertia 
as something really fundamental which is not so) 
introduced by such an approximation is within 1\% only \cite{Goychuk09}. 
Moreover,
in any realistic experimental setup with some observed viscoelastic
memory kernel exhibiting normally different complex patterns, 
see e.g. in Refs. \cite{Mason,Mizuno},
one can do another expansion, not assuming any precise scaling, but
rather simply fitting the experiment.
Our flexible methodology will work anyway and the embedding dimension
$D=N+2$  can typically be even smaller. For example, for all observed
cases of biological subdiffusion  (typically several time decades) 
the choice of $N=6$ is sufficient (will be published elsewhere).
We are interested here but in a faithful Markovian embedding of the
FBM type  of subdiffusion (which formally requires infinite dimensions) and 
integrate GLE using embedding dimension $D=18$ ($N=16$) which provides
excellent approximation.

\begin{figure}[t]
\centering
\includegraphics[width=8cm]{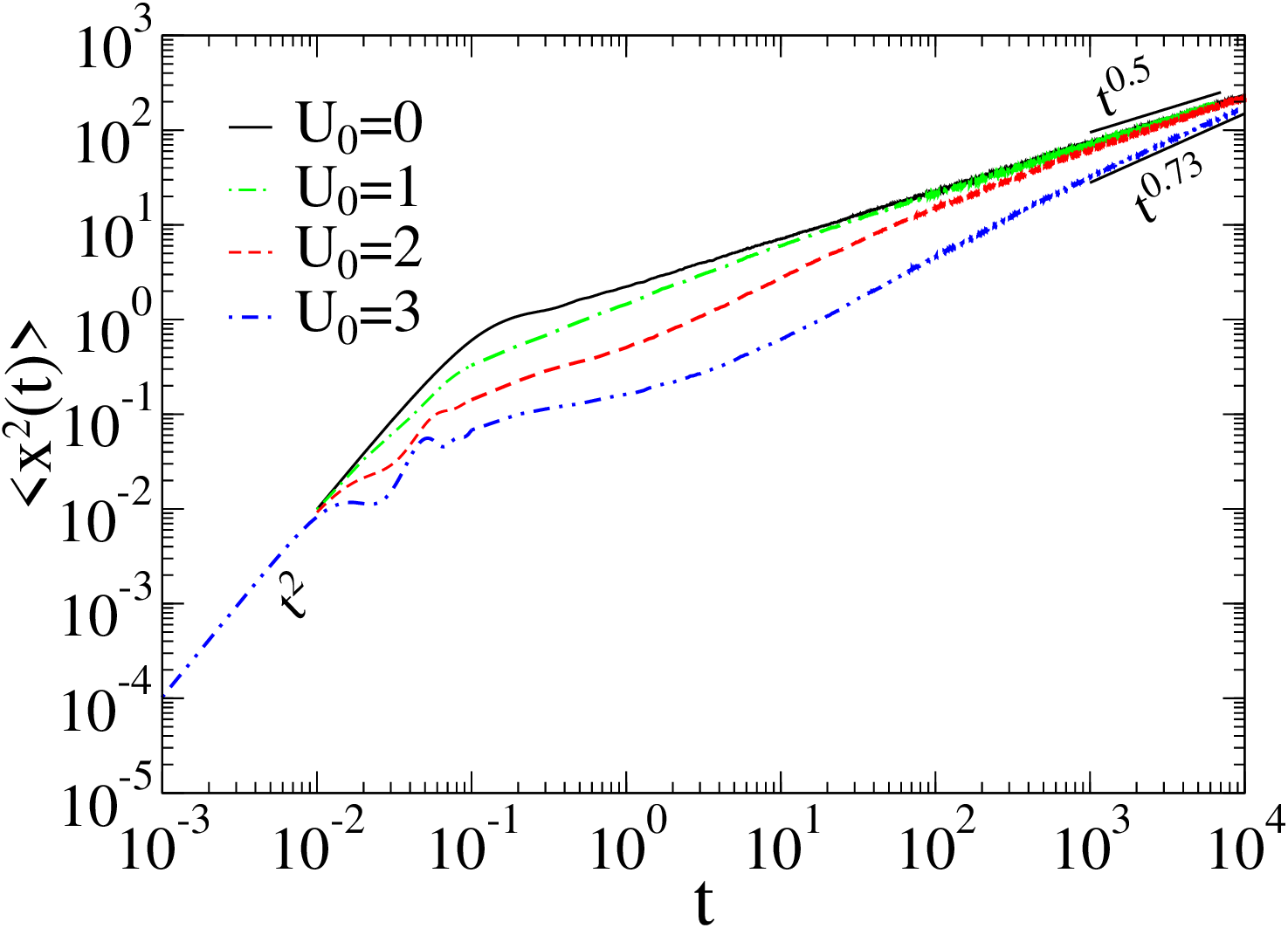}
\caption{(Color online) Stochastic simulations with embedding
dimension $D=18$ for different potential amplitudes (in units of $k_BT$).
The initial velocities are thermally distributed and 
initial diffusion is always ballistic. It changes into subdiffusion which
is asymptotically not sensitive to the barrier height. The transient
to this asymptotic regime can be however very slow, depending on the
barrier height. For $U_0=3$, it is still not completed, $\alpha_{\rm eff }(t)
\approx 0.73$ for $10^3<t<10^4$. It will but decrease to $\alpha=0.5$, because
of the subdiffusion in periodic potential cannot be faster than free subdiffusion,
even if we cannot arrive at this regime numerically within two weeks of simulation. 
Stochastic Heun algorithm is used with time step $\Delta t=10^{-4}$,
$\alpha=0.5$, $r=10$, and  $n=10^4$ trajectories for 
the ensemble averaging.
} 
\label{Fig2}
\end{figure}

\section{Results}

It is convenient to scale the coordinate $x$ in units of $L$, 
and the time in units of 
$\tau=[L^2\eta_{\alpha}/(k_BT)]^{1/\alpha}$, so that the particle
freely subdiffuses over a distance of the order $L$ in time  $\tau$.
More precisely, in this scaling $D_{\alpha}^{(0)}=1$ for the strict power 
law memory kernel  at any temperature. The potential amplitude
parameter $U_0$ is scaled in the units of $k_BT$ and the driving force
in the units of $k_BT/L$.   The frequency is scaled in the units of $\tau^{-1}$. 
Furthermore, the influence of the inertia effects is captured by the parameter 
$r=\tau/\tau_b$, where $\tau_b=L/v_T$ is the ballistic time 
for vanishing friction and
$v_T=\sqrt{k_B T/m}$ is the rms thermal velocity. High damping corresponds
to $r\gg 1$ \footnote{Another useful scaling is: time  in 
$\tau_v=(m/\eta_{\alpha})^{1/(2-\alpha)}$ and energy in
$E_0=L^2\eta_{\alpha}^{2/(2-\alpha)}/m^{\alpha/(2-\alpha)}$.
The temperature is scaled then as $\tilde T=k_BT/E_0$. The connection
between these two scalings is provided by: $\tau_v=\tau r^{-2/(2-\alpha)}$
and $\tilde T=r^{-2\alpha/(2-\alpha)}$.
One can easily rescale results at the fixed temperature (like
in this paper) from one scaling
to another one. Other scalings are also possible.}. We used the stochastic Heun algorithm \cite{Gard} to integrate
the system of stochastic differential equations (\ref{embedding}). It is of the
second order of weak convergency in the integration time step 
$\Delta t$ for our particular case of additive noise. The Mersenne Twister
pseudorandom number generator was used to produce the uniformly distributed random
numbers which were transformed into Gaussian random numbers in accordance with the
Box-Muller algorithm. With the time step $\Delta t=10^{-4}$ and $n=10^4$ trajectories
used for the ensemble averaging, the typical
accuracy of our simulations is within margin of several percents \cite{Goychuk09}, 
as tested by comparison with
the exact analytical results available for the potential-free subdiffusion and
for parabolic potentials. This is a very good quality
for stochastic numerics. It must be also noted that the use of double precision
floating-point arithmetics cannot be avoided to arrive at convergent results.

\subsection{ Static ratchet potential}

The influence of static ratchet 
potential $U(x)$ on the subdifussion  is illustrated in Fig. 
\ref{Fig2}. After a short ballistic stage (within a potential well), 
the diffusion  can look (depending on the potential barrier height)
initially  closer to normal. This is because of a finite mean residence time
in a potential well exists and the escape kinetics, being asymptotically
stretched-exponential, tends gradually to the normal exponential kinetics 
with an increase of the potential barrier height \cite{Goychuk09}.
However, its slows down and asymptotically, independently of the barrier
height (which is about  $2.2 U_0$ for static potential), 
reaches the boardline of free subdiffusion which clearly cannot be crossed.
This typical behavior can be characterized by a time-dependent exponent 
$\alpha_{\rm eff}(t)$ defined from the slope of 
$\langle \delta  x^2(t)\rangle$ curve in the double-logarithmic
coordinates: $\alpha_{\rm eff}\to \alpha$
with $t\to\infty$, independently of $U_0$. 
Such a behavior reflects the physical nature of viscoelastic  
subdiffusion which is due to the long-range correlations in the particle's
coordinate increments \cite{Goychuk09}, being in a sharp contrast with
the semi-Markov CTRW subdiffusion, where such correlations are absent 
in principle. These are not the rare events of the escape from potential
wells which determine asymptotically the temporal pace of diffusion, 
but the anti-persistent nature of viscoelastic subdiffusion which
finally wins and limits diffusion in the periodic potentials  by the 
free subdiffusion
limit which is finally attended regardless the potential height. 
The transient to this astounding asymptotical behavior 
is, however,
extremally slow and for this reason  may be not achieved in practice. 
Even if the presence of periodic potential does not 
influence subdiffusion asymptotically (and therefore the
adiabatically rocking subdiffusive ratchets are simply impossible), 
it does profoundly influence the whole time course of diffusion.
This fact, which seems to go beyond any analytical 
treatment, is at the heart of anomalous ratchet effect. 
This physical picture implies  
that the subdiffusive rectified
current has a resonance-like dependence against the driving frequency 
$\Omega$, because of in the limit $\Omega\to\infty$ the ratchet effect must 
{\it asymptotically} vanish as well. 

\begin{figure}[t]
\centering
\includegraphics[width=8cm]{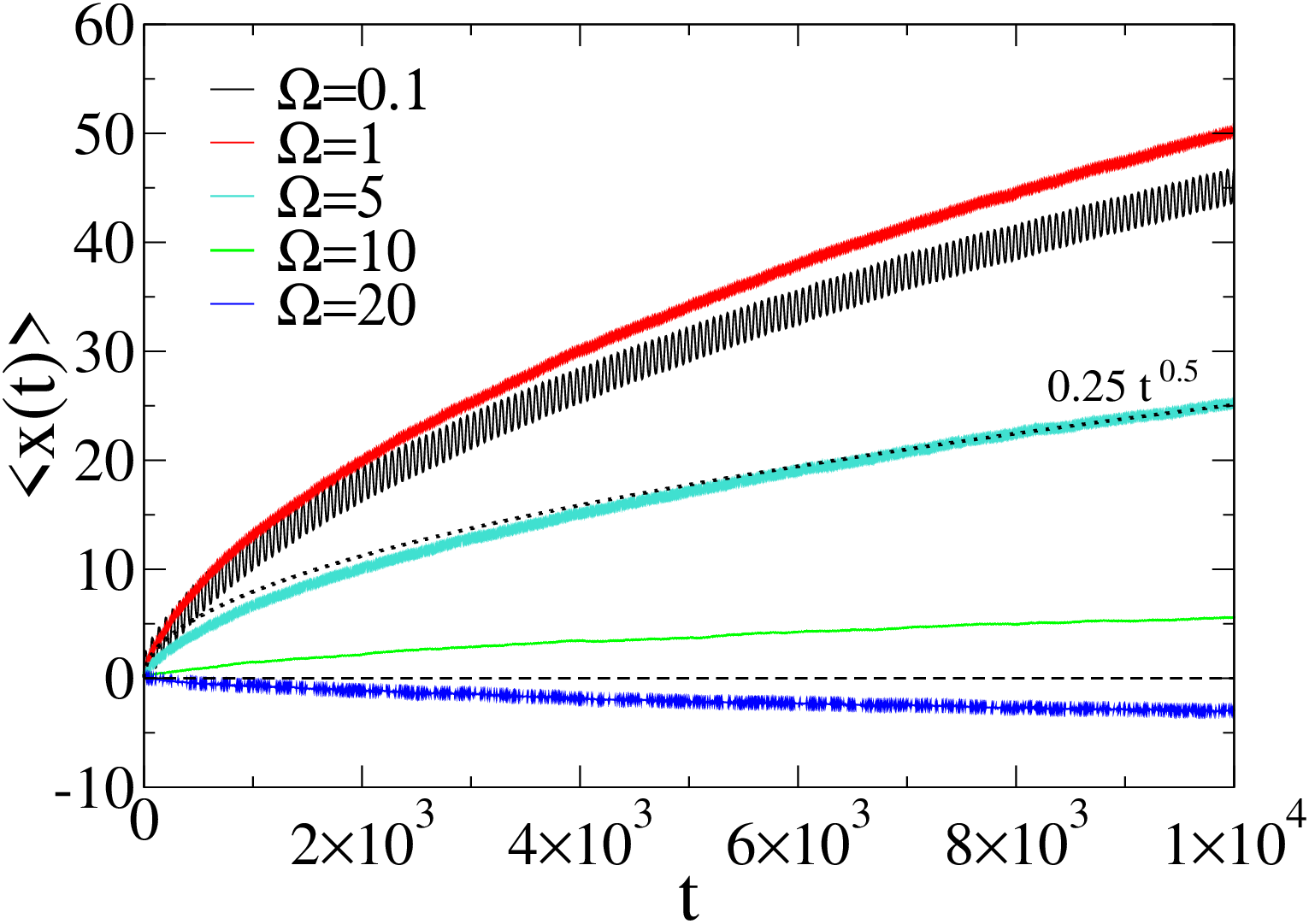}
\caption{(Color online) Mean particle position as function
of time for different driving frequencies,
$U_0=3$, $A=4$. Other parameters are the same as in Fig. \ref{Fig2}.
The subvelocity $v_{\alpha}$ is obtained by fitting the numerical dependences
by $v_{\alpha}t^{0.5}/\Gamma(1+\alpha)$ within the last window of length 
$\Delta {\cal T}=10^3$ in simulations (as it is shown for $\Omega=5$). 
Notice that this provides only an estimate
since still $\alpha_{\rm eff}(t)\approx 0.58$.
The smaller is $U_0$, the more reliable is this estimate, since $\alpha_{\rm eff}$
is better relaxed to the asymptotic value $\alpha=0.5$.
} 
\label{Fig3}
\end{figure}  

\subsection{ Rocking ratchets} 

\begin{figure}[t]
\centering
\includegraphics[width=8cm]{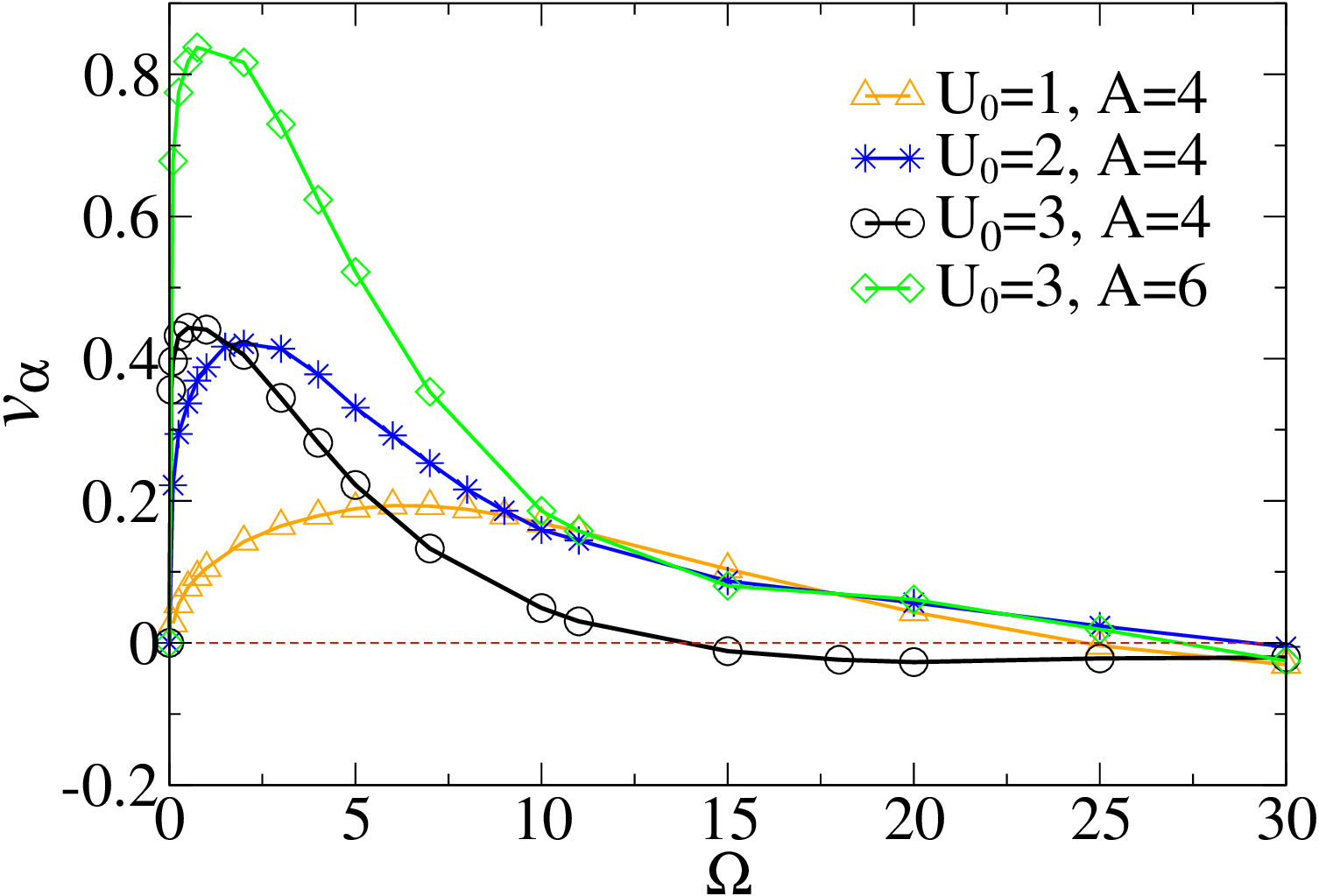}
\caption{(Color online) Anomalous current (subvelocity $v_{\alpha}$) as function of 
the driving frequency for different $U_0$ and $A$. 
Rectification vanishes for both adiabatically
slow and for a very fast driving. A current inversion is observed
for sufficiently large $\Omega$ in all cases.  }
\label{Fig4}
\end{figure}

To verify these qualitative considerations, 
we have tilted periodically the ratchet potential in  Eq.
(\ref{potential}) forth and back by $f(t)=A\cos(\Omega t)$. 
The amplitude $A$ is scaled in
the  units of $k_BT/L$. For a given $U_0$ (in units of
$k_BT$) there are two critical values of $A$ for maximal force $f(t)$:
(i)  $A_1=(3\pi/2) U_0\approx 4.71 U_0$ (when the potential
barrier vanishes  for the forward maximal tilt)
and (ii) $A_2=2A_1$ (same for the backward tilt).
Moreover, for $A$ between $A_3=\pi U_0$ and $A_1$, the potential $U(x)$ has
two small barriers within each spatial period of $V(x)$ for the forward tilt.
Generally, for
a subcritical driving the  potential barrier towards the direction of tilt 
is smaller for the forward tilt than for the backward tilt of equal amplitude.  
Therefore, intuitively the rectification
current should flow in the positive direction for $U_0>0$, cf. Fig. \ref{Fig1},
and this is indeed the fact, cf. Fig. \ref{Fig3}, for a sufficiently slow
driving.  However,
for the normal  diffusion ratchets the current inversion
is possible for a sufficiently high driving frequency  $\Omega$
\cite{Bartussek}. A similar counter-intuitive inversion was found 
also in present case, see in Figs. \ref{Fig3},\ref{Fig4}.  

Subdiffusive currents we define through the subvelocity 
\footnote{The limiting procedure should be understood here in a 
physical sense, i.e. $t$ is large but still
much smaller than the memory cutoff $\tau_c$. Otherwise, the 
proper mathematical limit will yield infinity for a memory kernel with cutoff.} 
\begin{eqnarray}
v_{\alpha}=\Gamma(1+\alpha) \lim_{t\to \infty}\frac{\langle
x(t)\rangle}{t^{\alpha}}.
\end{eqnarray}
Likewise, the subdiffusion coefficient in the potential is defined as
\begin{eqnarray}
D_{\alpha}=\frac{1}{2}\Gamma(1+\alpha) \lim_{t\to \infty}\frac{\langle \delta
x^2(t)\rangle}{t^{\alpha}},
\end{eqnarray}
so that it should coincide with $D_{\alpha}^{(0)}$ in the absence of potential. 
Furthermore, the quality of the anomalous rectification
can be characterized by a generalized Peclet number which we define
as  ${\rm Pe}_{\alpha}=v_{\alpha}L/D_{\alpha}$ by analogy with the
normal diffusion case \cite{Peclet}. 

It must be noted that practically we are dealing with 
some time-dependent $\alpha_{\rm eff}(t)$ in our numerics which relaxes very
slow (depending on $U_0$) to the asymptotic value $\alpha$. 
This asymptotic value is not easy to obtain for high $U_0$. 
To arrive at the end point $t=10^4$ in our simulations for one set of parameters
and to have numerically reliable convergent
results one has to propagate dynamics for about two weeks 
($\Delta t=10^{-4}, n=10^4$) on a single node
of our modern Linux cluster using a highly optimized for the ``number crunching''
FORTRAN Intel compiler. We derive the corresponding values of $v_{\alpha}$
and $D_{\alpha}$ by fitting the dependencies $\langle x(t)\rangle$ and
$\langle \delta x^2(t)\rangle $ with the $a t^{0.5}$
dependence (extracting the corresponding $a$) 
within the last time window of length 
$\Delta {\cal T}=10^3$ in the simulations.
For $U_0\leq 2$ this gives reliable results, and for $U_0=3$ the numerically
derived values are less reliable (see but a fitting
in Fig. \ref{Fig3} for $\Omega=5$ 
to get an idea on the quality of approximation). This is 
because of the corresponding $\alpha_{\rm eff}(t)$ is still in fact
about $0.58$ within this time window, if to consider $\alpha_{\rm eff}$ as an independent 
fitting parameter, i.e. it still did not relax 
completely to $\alpha$.  This value is but essentially lower and closer 
to $\alpha$  than the corresponding 
$\alpha_{\rm eff}\approx 0.73$ 
of the non-driven subdiffusion in the same potential, cf. Fig. \ref{Fig2}. 
Generally, periodic driving 
accelerates the ultraslow relaxation of $\alpha_{\rm eff}(t)$ to $\alpha$. However,
for $U_0=3$ it is, strictly speaking, still not quite achieved. We can compare only
asymptotic transport coefficients in different potentials (a comparison makes 
just a little
sense for some snapshot values $\alpha_{\rm eff}(t)$). For 
larger $U_0$ (e.g. $U_0=4$), it becomes
practically impossible to estimate these quantities. 
For this reason, such cases were not feasible for a
quantitative study, even though the anomalous ratchet effect exists of course also.  
This striking feature of a time-dependent $\alpha_{\rm eff}(t)$
must be kept in mind when to study subdiffusive ratchets experimentally.

The absence  of {\it asymptotic} ratchet effect in adiabatic driving limit 
(a {\it transient} one
yet exists!) is in a 
sharp contrast with the normal diffusion rocking ratchets
where the current is maximized  for $\Omega\to 0$ \cite{Bartussek,Reimann}. 
An inspection of single particle trajectories (not shown)
indicates that the optimal frequency in our case corresponds to a
synchronization of the potential tilts and the particle motion between potential
wells being a kind of stochastic resonance \cite{SR}.
In this respect, our anomalous rocking ratchet resembles more
normal flashing ratchet than normal rocking ratchet.
Indeed, if during a driving (half-)period the particles can move over
many lattice periods  the potential profile
asymptotically does not matter and the current is suppressed. For the frequencies
much larger than this inverse characteristic time the current is again suppressed
(here per an analogy with normal diffusion ratchets: too frequent force 
alternations hamper the motion). This explains the occurrence
of the resonance like feature. The maximal amplitude of the rectification current
first increases with increasing the barriers 
height, see in Fig. \ref{Fig4}, where the current 
is maximal for the fixed $A$ at the highest potential barrier
being optimized for $\Omega$.
It will however become obviously 
suppressed with a further increase of $U_0$.
There are optimal potential amplitudes depending
on driving strength and frequency, just like in the case of normal ratchets.
The unexpected analogy to the flashing normal ratchets is, however, not complete
as the (sub-)current inversion for sufficiently high frequency $\Omega$
shows.  This feature is similar to one in rocking normal ratchets and it
unfortunately cannot be explained in simple intuitive terms.

\begin{figure}[t]
\centering
\includegraphics[width=8cm]{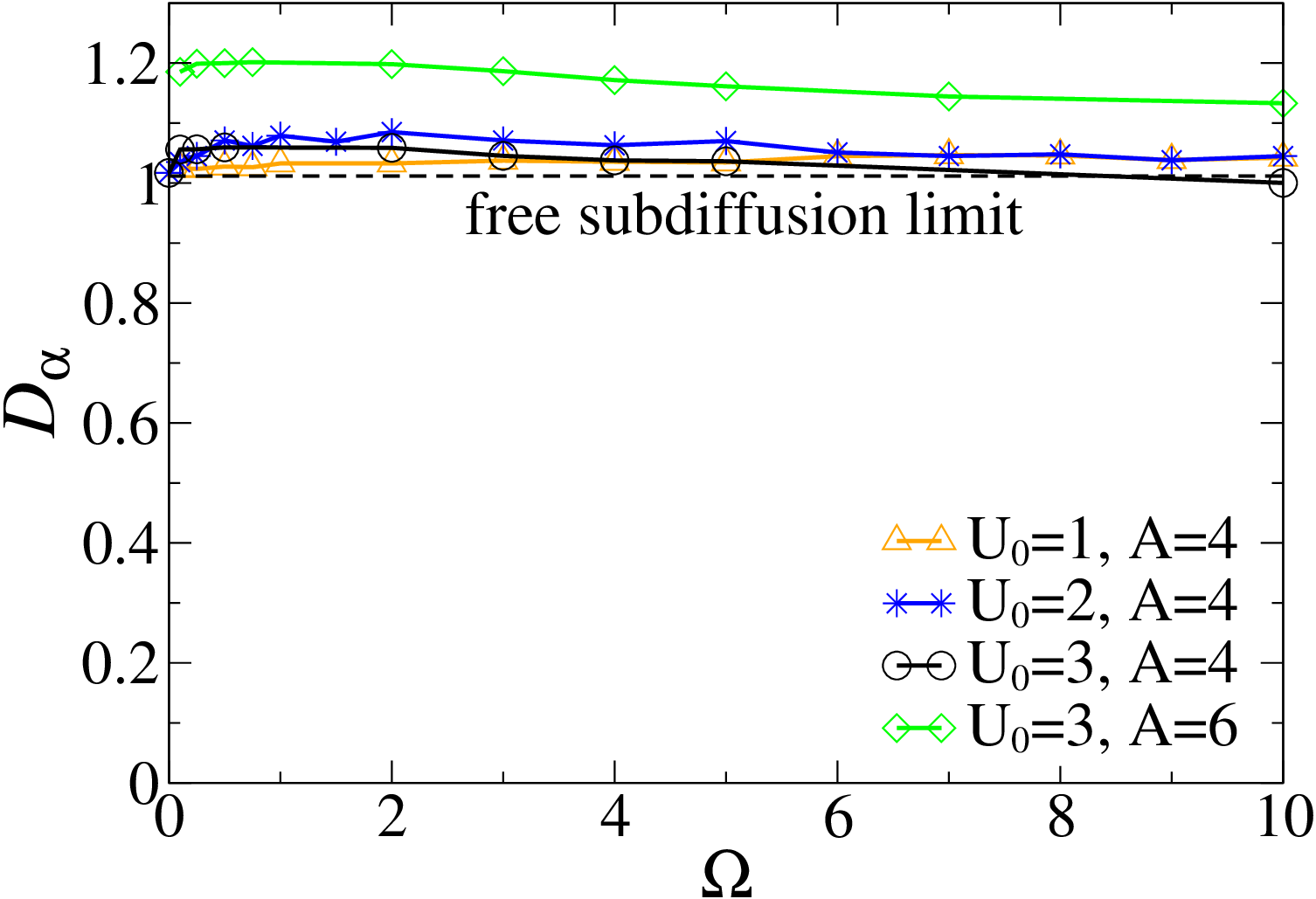}
\caption{(Color online) Subdiffusion coefficient as function of 
driving frequency for different $U_0$ and $A$. 
 }
\label{Fig5}
\vspace{1cm}
\end{figure}

The anomalous diffusion coefficient does not
display such a profound dependence on frequency as $v_{\alpha}$.
It becomes somewhat increased (less than few percent for $A=4$ and
no more than 20\% for $A=6$) 
as compare  
with the free subdiffusion coefficient 
(whose numerical value $D_{\alpha} \approx 1.012$ agrees well with 
the theoretical value of $D_{\alpha}^{(0)}=1$ in the used scaling).
However, this effect is relatively small, compare Fig. \ref{Fig4}
and Fig. \ref{Fig5}. 
This remarkable feature is of the same
origin as insensitivity of the GLE subdiffusion to the height of
periodic potential. It is in a sharp contrast
with the normal diffusion case.
For this reason, the dependence of the
generalized Peclet number on frequency just resembles the behavior of 
the absolute value
of subcurrent in Fig. \ref{Fig4}.

\begin{figure}[t]
\centering
\includegraphics[width=8cm]{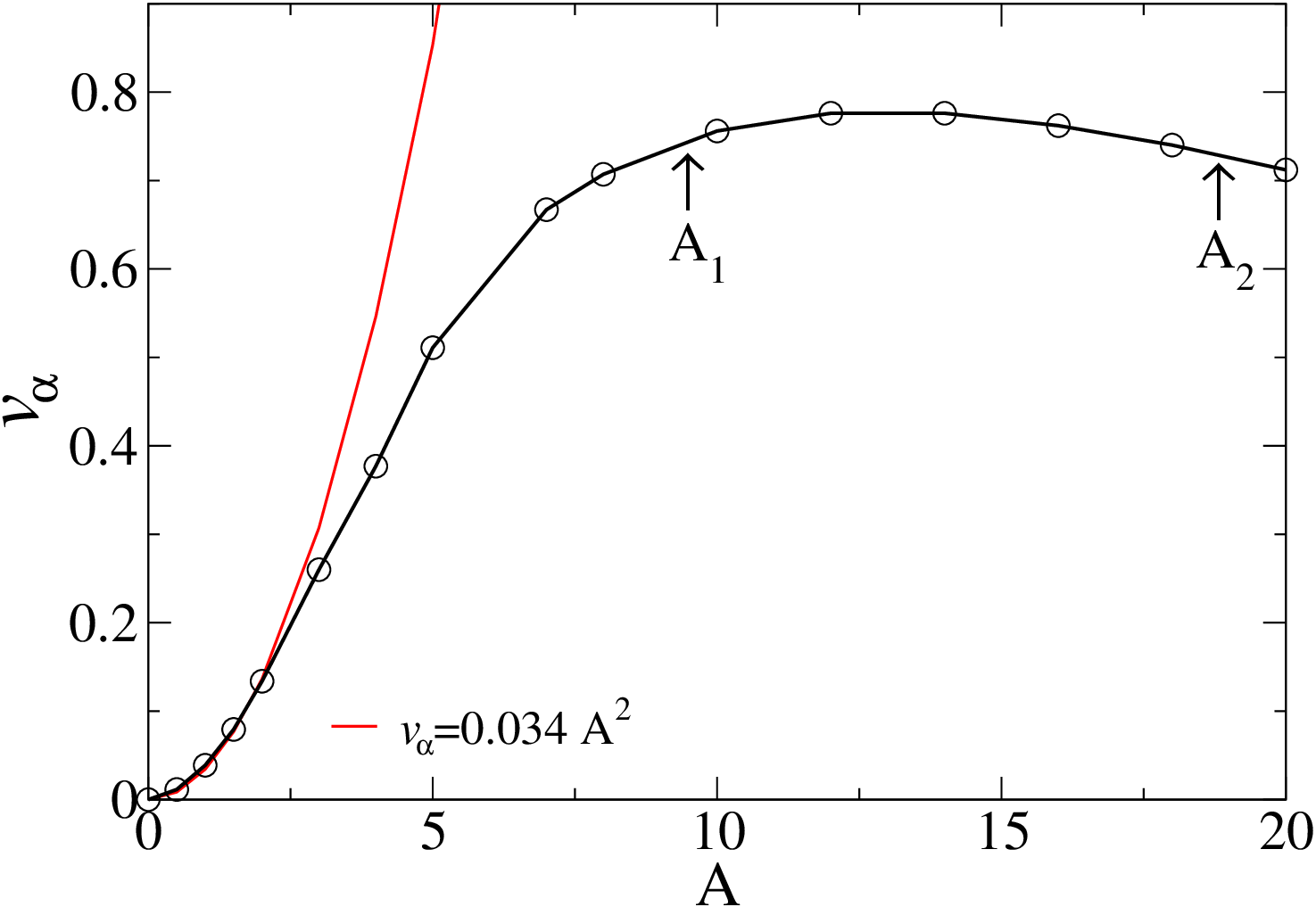}
\caption{(Color online) Anomalous current as function of 
driving amplitude. Quadratic response fit is also displayed for
comparison. $U_0=2$, $\Omega=1$. }
\label{Fig6}
\end{figure}
The dependence of rectification current on the driving field amplitude $A$
is shown in Fig. \ref{Fig5} for $U_0=2$ and $\Omega=1$.
It displays the same quadratic nonlinear 
dependence
on driving amplitude for a small driving strength as in normal
ratchets \cite{Bartussek,Reimann}. Indeed, the rectification current cannot be present
within the linear response (due to the Onsager symmetry of 
kinetic coefficients). It first emerges as the lowest, second order nonlinear
effect which is not forbidden by symmetry considerations.  
Moreover, beyond perturbation theory 
the rectification current  displays a very broad maximum for
$A_1< A < A_2$, in accord with intuition.

\section{Discussion and Summary}

In this work we put forward a model of subdiffusive Brownian ratchets within
the GLE description. Notably, a strict power law kernel is not required to
reproduce subdiffusion on a very long time scale (the theoretical asymptotics
of normal diffusion for $t>10^{11}$ cannot be even reached numerically
on the available computers what makes it practically irrelevant in our simulations), and 
the Markovian embedding
dimension can be surprisingly small, allowing to accurately 
approximate nonlinear subdiffusive dynamics driven by Gaussian $1/f^{1-\alpha}$ 
noise.  Moreover, 
power law memory kernels with cutoffs are in fact more physical than
strict power laws and their expansion over exponentials can reflect the
corresponding relaxation spectrum of viscoelastic response of the surrounding medium.

The way we did this Markovian embedding, i.e.  approximating the 
fractional Gaussian noise by a sum of Ornstein-Uhlenbeck processes
is also insightful. In particular, it allows to justify the view of the escape
dynamics out of a potential well as a rate process with non-Markovian fluctuating rate, when
the potential barrier exceeds $k_BT$ \cite{Goychuk09}. Such an escape dynamics
is asymptotically stretched exponential, but it tends gradually to a 
single-exponential with the increase of the potential barrier.
Then the rate fluctuations
gradually vanish and for sufficiently high barriers (exceeding e.g. $12\;k_BT$ for $\alpha=0.5$, 
and about
$9\;k_BT$ for $\alpha=0.75$) the escape dynamics is excellently approximated by
the celebrated non-Markovian rate expression \cite{HTB90,Grote,HanggiMojtabai,Pollak}, 
as it was shown by us
recently \cite{Goychuk09}. However, even for such high potential barriers, when the
escape dynamics out of a potential well becomes practically exponential, the diffusion 
in a periodic potential is not normal, but anomalously slow. This is because of 
viscoelastic subdiffusion is based not on the anomaly of the residence time
distributions (i.e. divergent, or extremally large mean residence time, like in the case of 
CTRW subdiffusion and akin mechanisms), but on the long-time
anti-correlations in the particle displacements (and velocity alternations).
This fact in combination with ergodicity  makes GLE subdiffusion physically much more appealing
scenario, especially in biological applications where it can be combined with the quenched
disorder due to the medium's inhomogeneity (not necessarily leading alone to 
the emergence of subdiffusion).

The subdiffusion in periodic potentials is described by some 
time-dependent $\alpha_{\rm eff}$ which relaxes very slowly  to $\alpha$. The time pace
of this relaxation depends very essential on the potential amplitude \cite{Goychuk09}, even
if asymptotically this subdiffusion is not sensitive to the presence of potential \cite{Chen}
because of the  antipersistent, sluggish  nature of this subdiffusion finally wins over 
thermally assisted hops between potential wells. The latter ones 
can take place quite frequently and do not provide
a transport limiting step asymptotically (a huge difference with the CTRW approach!). 
This circumstance makes asymptotic ratchet effect impossible in the limit of
adiabatically slow driving with vanishing driving frequency, $\Omega\to 0$. 
A somewhat similar suppression of the response to adiabatically slow driving was observed 
also in the non-Markovian stochastic resonance \cite{GoychukHanggi03,GoychukHanggi04}.
It seems to be a general feature of the non-Markovian dynamics with infinite memory which
culminates in the death of linear (and not only!) response in the case of subdiffusive
CTRW and the associated fractional Fokker-Planck dynamics \cite{Sokolov06,PRL07,Heinsalu09}. 
However, for a finite $\Omega$ the asymptotic anomalous current response of  subdiffusive
ratchet dynamics is present. In the lowest order of driving strength $A$ it emerges
as a nonlinear quadratic response to driving which violates the symmetry of 
thermal detailed balance  in the absence of driving, 
for the unbiased potential.
This is similar to the out of equilibrium physics of 
normal diffusion ratchets. The rectification current is optimized
not only for a driving strength between two critical values 
$A_1=3\pi U_0/(2 L)$ and 
$A_2=2A_1$ for the consider potential, but also versus $\Omega$ and $U_0$ for the fixed
temperature $T$. This is because of the rectification current vanishes in all these limits:
$\Omega\to 0$, $\Omega\to\infty$, $U_0\to 0$, $U_0\to \infty$. The visual inspection of 
the single particle trajectories shows that the maximum of subvelocity versus
$\Omega$ corresponds to an optimal synchronization of the particle motions between potential
wells and the potential tilts. This mechanism is somewhat similar to the current
optimization for flashing, or pulsing normal diffusion ratchets \cite{Hanggi09}.
However, the physics is different. In our case, the effect results 
because of the potential is not felt strongly by particles
traveling large distances between the potential alternations. Moreover, for sufficiently
large frequencies particles subdiffuse in the counter-intuitive direction, 
i.e. the anomalous current inversion occurs, cf. Figs. \ref{Fig3}, \ref{Fig4}.
Like in the case of normal diffusion rocking ratchets \cite{Bartussek}, this effect
does not have a simple intuitive explanation.  

To conclude, this work put forward a generalization of the pioneering contributions
\cite{Magnasco,Bartussek} to the realm of subdiffusive Brownian ratchets 
in viscoelastic media rocked by a
periodic force. The author is confident that the bulk of future research on 
subdiffusive Brownian ratchets
is ahead because of their surprising and counter-intuitive features which call for experimental
verification.


\end{document}